\documentclass[pdflatex,sn-mathphys-num]{sn-jnl}

\usepackage{graphicx}%
\usepackage{multirow}%
\usepackage{amsmath,amssymb,amsfonts}%
\usepackage{amsthm}%
\usepackage{mathrsfs}%
\usepackage[title]{appendix}%
\usepackage{xcolor}%
\usepackage{textcomp}%
\usepackage{manyfoot}%
\usepackage{siunitx}
\usepackage{booktabs}%
\usepackage{algorithm}%
\usepackage{algorithmicx}%
\usepackage{algpseudocode}%
\usepackage{listings}%
\usepackage{ulem}

\theoremstyle{thmstyleone}%
\theoremstyle{thmstyletwo}%

\theoremstyle{thmstylethree}%

\begin{document}

\title[Article Title]{
Multiple charge transfer driven complex reaction dynamics: covalent bonding meets van der Waals interactions
}


\author[1,2]{\fnm{Ruichao} \sur{Dong}}
\equalcont{These authors contributed equally to this work.}
\author[3]{\fnm{Xiaoqing} \sur{Hu}}
\equalcont{These authors contributed equally to this work.}
\author[4]{\fnm{Owen Dennis} \sur{McGinnis}}
\author[1]{\fnm{Xincheng} \sur{Wang}}
\author[1]{\fnm{Yikang} \sur{Zhang}}
\author[1]{\fnm{Ahai} \sur{Chen}}
\author[4]{\fnm{Andreas} \sur{Pier}}
\author[4]{\fnm{Alexander} \sur{Tsertsvadze}}
\author[1]{\fnm{Huanyu} \sur{Ma}}
\author[1]{\fnm{Jinze} \sur{Feng}}
\author[4]{\fnm{Jessica} \sur{Weiherer}}
\author[4]{\fnm{Laura} \sur{Sommerlad}}
\author[4]{\fnm{Madeleine} \sur{Schmidt}}
\author[4]{\fnm{Niklas} \sur{Melzer}}
\author[4]{\fnm{Noah} \sur{Kraft}}
\author[4]{\fnm{Sina Marie} \sur{Jacob}}
\author[1]{\fnm{Zhenjie} \sur{Shen}}
\author[5]{\fnm{Noelle} \sur{Walsh}}
\author[3]{\fnm{Jianguo} \sur{Wang}}
\author[4]{\fnm{Reinhard} \sur{Dörner}}
\author[1,7]{\fnm{Kiyoshi} \sur{Ueda}}
\author*[3]{\fnm{Yong} \sur{Wu}}\email{wu$_{-}$yong@iapcm.ac.cn}
\author*[6]{\fnm{Florian} \sur{Trinter}}\email{trinter@fhi-berlin.mpg.de}
\author*[4,8]{\fnm{Till} \sur{Jahnke}}\email{jahnke@atom.uni-frankfurt.de}
\author*[1,2,9]{\fnm{Yuhai} \sur{Jiang}}\email{jiangyh3@shanghaitech.edu.cn}

\affil[1]{\orgdiv{Center for Transformative Science and School of Physical Science and Technology}, \orgname{ShanghaiTech University}, \orgaddress{\street{No. 393 Middle Huaxia Road}, \city{Shanghai}, \postcode{201210}, \country{China}}}

\affil[2]{\orgdiv{Shanghai Advanced Research Institute}, \orgname{Chinese Academy of Sciences}, \orgaddress{\street{No. 99 Haike Road}, \city{Shanghai}, \postcode{201210}, \country{China}}}

\affil[3]{\orgdiv{National Key Laboratory of Computational Physics}, \orgname{Institute of Applied Physics and Computational Mathematics}, \orgaddress{\street{No. 2 Fenghao East Road}, \city{Beijing}, \postcode{100088}, \country{China}}}

\affil[4]{\orgdiv{Institut für Kernphysik}, \orgname{Goethe-Universität Frankfurt}, \orgaddress{\street{Max-von-Laue-Str. 1}, \city{ Frankfurt am Main}, \postcode{60438}, \country{Germany}}}

\affil[5]{\orgdiv{MAX IV Laboratory}, \orgname{Lund University}, \orgaddress{\street{Box 118}, \city{Lund}, \postcode{22100}, \country{Sweden}}}

\affil[6]{\orgdiv{Molecular Physics}, \orgname{Fritz-Haber-Institut der Max-Planck-Gesellschaft}, \orgaddress{\street{Faradayweg 4-6}, \city{Berlin}, \postcode{14195}, \country{Germany}}}

\affil[7]{\orgdiv{Department of Chemistry}, \orgname{Tohoku University}, \orgaddress{\street{No. 41 Kawauchi}, \city{Sendai}, \postcode{980-8578}, \country{Japan}}}

\affil[8]{\orgname{Max-Planck-Institut für Kernphysik}, \orgaddress{\street{Saupfercheckweg 1}, \city{Heidelberg}, \postcode{69117}, \country{Germany}}}

\affil[9]{\orgdiv{School of Physics}, \orgname{Henan Normal University}, \orgaddress{\street{No. 46 Jianshe East Road}, \city{Xinxiang}, \postcode{453007}, \state{Henan}, \country{China}}}




\abstract{Ultrafast charge transfer (CT) processes redistribute electronic charge within and between molecular units and play a central role in many physical, chemical, and biological phenomena. However, the microscopic pathways of multiple CT events, including the coupled structural evolution and energy redistribution, are challenging to disentangle experimentally in complex systems. To obtain controlled insight into such dynamics, well-defined properties are required. Here, we investigate the N$_2$Ar dimer, which combines a covalent bond with a weak van der Waals interaction, using site-selective synchrotron photoionization and coincident detection of electrons and ions. Combined with {\it ab initio} calculations, this approach enables step-by-step tracking of ultrafast CT and fragmentation dynamics. We find that the dimer's structural evolution triggers a second CT event, opening complex reaction pathways in which electrons are transferred back and forth between Ar and N$_2$, through two nonadiabatic transitions involving conical intersections. These results demonstrate that sequential multiple CT-induced transitions, even in a simple dimer, provide controlled insight into nonadiabatic reaction mechanisms relevant to complex systems.}

\maketitle

Ultrafast charge transfer (CT), which redistributes electronic charge within or between molecular units, is a fundamental process underlying functionality in a wide range of natural and engineered systems \cite{RevModPhys.65.599, doi:10.1126/science.1253607, PhysRevLett.113.073001, article_Boll, articleZhangNChe,
article_Gopakumar, articleCTTS}. 
Charge and energy transport span length scales from small dimers and isolated molecules in the gas or liquid phase to photoactive materials and proteins \cite{10.1063/1.4996505, D2CS00006G, article_OSC, annurevETProt, article_OSC2021, KAUFMAN2024116546}. Tracing complex reactions induced by multiple CT via bulk ensemble measurements is often challenging, and key microscopic aspects of these processes can remain difficult to resolve. Access to well-defined molecular systems therefore provides an important route to uncover transient stages of CT dynamics and to elucidate the elementary quantum motion of electrons and nuclei involved in these processes \cite{doi:10.1126/science.abj3007, doi:10.1126/science.ads4369}.

In practice, charge and energy exchange typically occur in complex environments where covalent interactions coexist with weaker noncovalent forces such as van der Waals bonding.
The latter, despite being the weakest intermolecular interaction, plays a crucial role in determining the structure and dynamics of molecular aggregates, condensed phases, and biological matter \cite{articleDelrio, PhysRevLett.107.185701, doi:10.1073/pnas.1208121109, doi:10.1126/science.1254132,  10.1063/1674-0068/cjcp1901007}.
Simple rare-gas dimers, however, are not ideal systems for studying complex multiple CT-driven reaction dynamics, as they usually dissociate promptly after a single CT event \cite{PhysRevLett.105.263202, article_Ren}.
A suitable prototype system is the N$_2$Ar dimer, which features a strong covalent bond within N$_2$ and a weak van der Waals bond between N$_2$ and Ar. Its distinct properties make it a prototypical emulating model for experiment and quantum-chemical calculations to probe reaction mechanisms relevant to complex systems. First, the nearly matching ionization potentials \cite{10.1063/1.4750980, PhysRevLett.111.083003} of its two constituents make intermolecular reactions more effective under photoionization and photoexcitation scenarios by creating numerous crossings on the potential energy surfaces.
Second, its structural flexibility, especially the stretching of its bonds \cite{doi:10.1126/science.1074201} enables access to a manifold of nonadiabatic pathways through conical intersections and initiates the occurrence of multiple CT. Therefore, a system as simple as N$_2$Ar can effectively mimic the ensemble reactions occurring in complex condensed chemical and biological systems.

In more detail, in this single-molecule mixed-interaction system, it is possible to study multiple CT-induced bond cleavage and formation, novel reaction channels, and environmental reactions. Conversely, already a single CT (SCT) is a nontrivial process in this loosely bound entity: the transferred charge preferentially localizes at the Ar site rather than at the N$_2$, a topic studied for almost half a century \cite{articleZhangNChe, NIKITIN1987313, doi:10.1126/science.1074201, 10.1063/1.459179, 10.1063/1.1413980, doi:10.1021/acs.jpclett.1c00798}. More recently, the formation of [NAr]$^+$ was attributed to tunneling of a heavy N$^+$ ion along the reaction [N$_2$Ar]$^{2+}\rightarrow ~$N$^+$ + [NAr]$^+$ \cite{article_NAr}. When [N$_2$Ar]$^{2+}$ is created in collisions with highly charged ions, the double charge is expected to localize at the N$_2$ site, raising questions about the interplay between CT and competing processes such as interatomic Coulombic decay. These surprising findings have sparked growing interest, but their underlying mechanisms remain unresolved.

X-ray-induced inner- or core-shell ionization offers an element-specific route to prepare this system in well-defined charge states. In this work, we investigate experimentally and theoretically the photoionization-induced nonadiabatic dynamics in N$_2$Ar, focusing on ultrafast multiple CT processes. Using datasets recorded at different photon energies, we selectively initiate double ionization (DI) of either N$_2$ or Ar via Auger decay or direct double ionization, thereby preparing well-defined initial states. By analyzing the momenta of dimer fragments and electrons in the final state, we have unambiguously disentangled various CT pathways, enabling reconstruction of fragmentation pathways driven by SCT and double CT (DCT). Our findings demonstrate that even in such a simple dimer multiple CT-induced transitions can drive complex reaction channels, providing fundamental insight into charge redistribution as a core nonadiabatic mechanism underlying functional processes in complex natural systems.

\section*{Expected Charge Transfer Channels and Involved Potential Energy Surfaces}\label{sec2}

To understand how the observed reaction channels and ultrafast CTs evolve from the doubly charged N$_2$Ar dimer, we calculated the corresponding potential energy surfaces (PESs) and traced possible fragmentation pathways. Details of the calculations are in the Methods section. For [N$_2$Ar]$^{2+}$, charge localization can occur in three configurations: N$_2^{2+}$Ar, N$_2$Ar$^{2+}$, and N$_2^{+}$Ar$^{+}$. The high density of electronic states and the vibrational/rotational degrees of freedom of N$_2$ lead to significant overlap among the PESs over a wide range of N$_2$Ar bond lengths and angles. This extensive overlap facilitates the exceptionally rapid nonadiabatic CT dynamics, which is rare in simpler systems like Ar$_2$ dimers. Selective ionization using X-ray photons prepares specific initial states: Auger decay following removal of an N~1s or Ar~2p electron generates N$_2^{2+}$Ar (e.g., $\mathrm{X^1}\,\Sigma_g^+$, a$^3\Pi_u$, $\mathrm{D^1}\,\Sigma_u^+$) or N$_2$Ar$^{2+}$ (e.g., $\mathrm{^3P}$, $\mathrm{^1D}$, $\mathrm{^1S}$) states, respectively \cite{Lundqvist_1996, LOWerme_1973}. Intermolecular interactions then drive the cluster ion's structural evolution, potentially leading to a nearly linear N$_2$Ar geometry.

\begin{figure}[ht]
\centering
\includegraphics[width=1.0\textwidth]{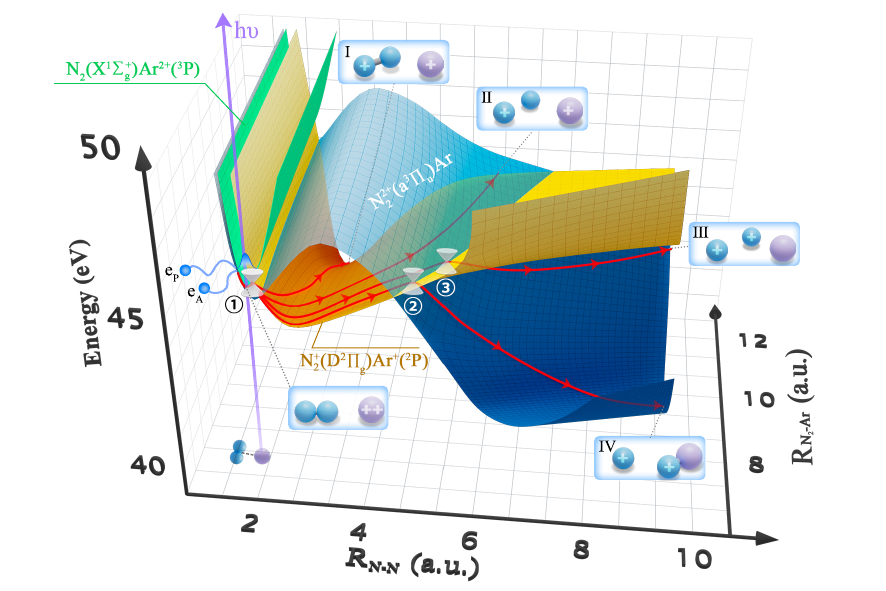}
\caption{\textbf{Schematic illustration of charge transfer processes from calculated potential energy surfaces (PESs).} Calculated PESs of the initial Auger final states, N$_2(\mathrm{X^1}\,\Sigma_g^+)$Ar$^{2+}$($\mathrm{^3P}$) (green) and N$_2^{2+}$(a$^3\Pi_u$)Ar (blue), produced following photoionization of the Ar~2p or N~1s shell, respectively. Ion pairs I (N$_2$$^{+}$, Ar$^{+}$) and II (N$^{+}$, Ar$^{+}$, N) are generated via conical intersections between these initial states and the N$_2^{+}$(D$^2\Pi_g$)Ar$^{+}$($\mathrm{^2P}$) (yellow) state. Subsequently, the ion pairs III (N$^{+}$, N$^{+}$, Ar) and IV (N$^{+}$, [NAr]$^{+}$) emerge through two additional conical intersections involving the N$_2^{+}$(D$^2\Pi_g$)Ar$^{+}$($\mathrm{^2P}$) and N$_2^{2+}$(a$^3\Pi_u$)Ar states. For clarity, the schematics highlight only the ionization pathways initiated at the Ar site.}\label{fig1}
\end{figure}


Figure~\ref{fig1} illustrates the two-dimensional PESs for three electronic states of the linear [N$_2$Ar]$^{2+}$ ion: N$_2^{2+}$(a$^3\Pi_u$)Ar, N$_2(\mathrm{X^1}\,\Sigma_g^+)$Ar$^{2+}$($\mathrm{^3P}$), and N$_2^{+}$(D$^2\Pi_g$)Ar$^{+}$($\mathrm{^2P}$). When the system starts in N$_2^{2+}$(a$^3\Pi_u$)Ar or N$_2$Ar$^{2+}$($\mathrm{^3P}$), the N-N vibrational wave packet lies within the crossing region with the N$_2^{+}$(D$^2\Pi_g$)Ar$^{+}$($\mathrm{^2P}$) surface, enabling prompt SCT to form the N$_2^{+}$ + Ar$^{+}$ ion pair. The N-N bond elongates because the equilibrium bond length of N$_2^{+}$(D$^2\Pi_g$) is much larger than that of neutral N$_2$ or N$_2^{2+}$(a$^3\Pi_u$). Coulomb repulsion then stretches the N$_2$-Ar bond, promoting dissociation along two distinct SCT pathways:
\\


\medskip
\hspace {1cm} N$_2$Ar$^{2+}$ / N$_2^{2+}$Ar $\xrightarrow{\text{CT}}$ N$_2^{+}$Ar$^{+}$ $\rightarrow$ N$_2^{+}$ + Ar$^{+}$ \hfill (I) \hspace{1.1cm}
\medskip

\medskip
\hspace {1cm} N$_2$Ar$^{2+}$ / N$_2^{2+}$Ar $\xrightarrow{\text{CT}}$ N$_2^{+}$Ar$^{+}$ $\rightarrow$ N$^{+}$ + N + Ar$^{+}$ \hfill (II) \hspace{1cm}
\medskip

As the N-N bond continues to elongate, the N$_2^{+}$(D$^2\Pi_g$)Ar$^{+}$($\mathrm{^2P}$) and N$_2^{2+}$(a$^3\Pi_u$)Ar PESs intersect again, allowing a second CT event. For clusters starting in N$_2^{2+}$(a$^3\Pi_u$)Ar, dissociation is two-step: N$_2^{2+}$(a$^3\Pi_u$) captures an electron from Ar, forming N$_2^{+}$Ar$^{+}$, and then, as the N-N bond stretches, an electron transfers back to Ar$^{+}$. If the initial state is N$_2$Ar$^{2+}$($\mathrm{^3P}$), neutral N$_2$ sequentially accepts two electrons from Ar$^{2+}$, ultimately driving dissociation. Crucially, if the N$_2$-Ar distance is sufficiently short after the second CT, the emerging N$^{+}$ ion can bind to neutral Ar, yielding the new [NAr]$^{+}$ species via ion transfer. This secondary surface crossing opens two additional dissociation pathways:
\\


\medskip
\hspace {1cm} N$_2$Ar$^{2+}$ $\xrightarrow{\text{CT}}$ N$_2^{+}$Ar$^{+}$ $\xrightarrow{\text{CT}}$ N$_2^{2+}$Ar $\rightarrow$ N$^{+}$ + N$^{+}$ + Ar \hfill (III) \hspace{1cm}
\medskip

\medskip
\hspace {1cm} N$_2$Ar$^{2+}$ $\xrightarrow{\text{CT}}$ N$_2^{+}$Ar$^{+}$ $\xrightarrow{\text{CT}}$ N$_2^{2+}$Ar $\rightarrow$ N$^{+}$ + [NAr]$^{+}$ \hfill (IV) \hspace{1cm}
\medskip

As we show below, these fragmentation channels (III and IV) occur exclusively following photoionization of the Ar~2p shell. While Fig.~\ref{fig1} shows crossings among three representative states, {\it ab initio} calculations reveal dozens of intersecting states, providing multiple SCT and DCT pathways.

\section*{Experimental Results and Discussion}\label{sec3}


In the following, we present the data recorded in two experimental campaigns, which were performed at the soft X-ray beamline P04 of PETRA~III and the Flexible Photo\-Electron Spectroscopy (FlexPES) beamline of MAX~IV. We used a COLTRIMS reaction microscope \cite{DORNER200095, JAHNKE2004229} for a coincident measurement of the momenta of electrons and ions generated after the absorption of the synchrotron light (see Materials and Methods for details). As indicated above, the doubly charged cluster ions, N$_2$Ar$^{2+}$ or N$_2^{2+}$Ar, were produced via Auger decay following site-selective X-ray photoionization of the Ar~2p (260~eV) and N~1s (420~eV) shells, respectively. In addition, direct DI leading to N$_2^{2+}$Ar was investigated at a photon energy of 43.1~eV. At this photon energy, a direct DI at the Ar site is energetically forbidden \cite{ELAND2003171, PhysRevLett.96.243402}.

\begin{figure}[h]
\centering
\includegraphics[width=1.0\textwidth]{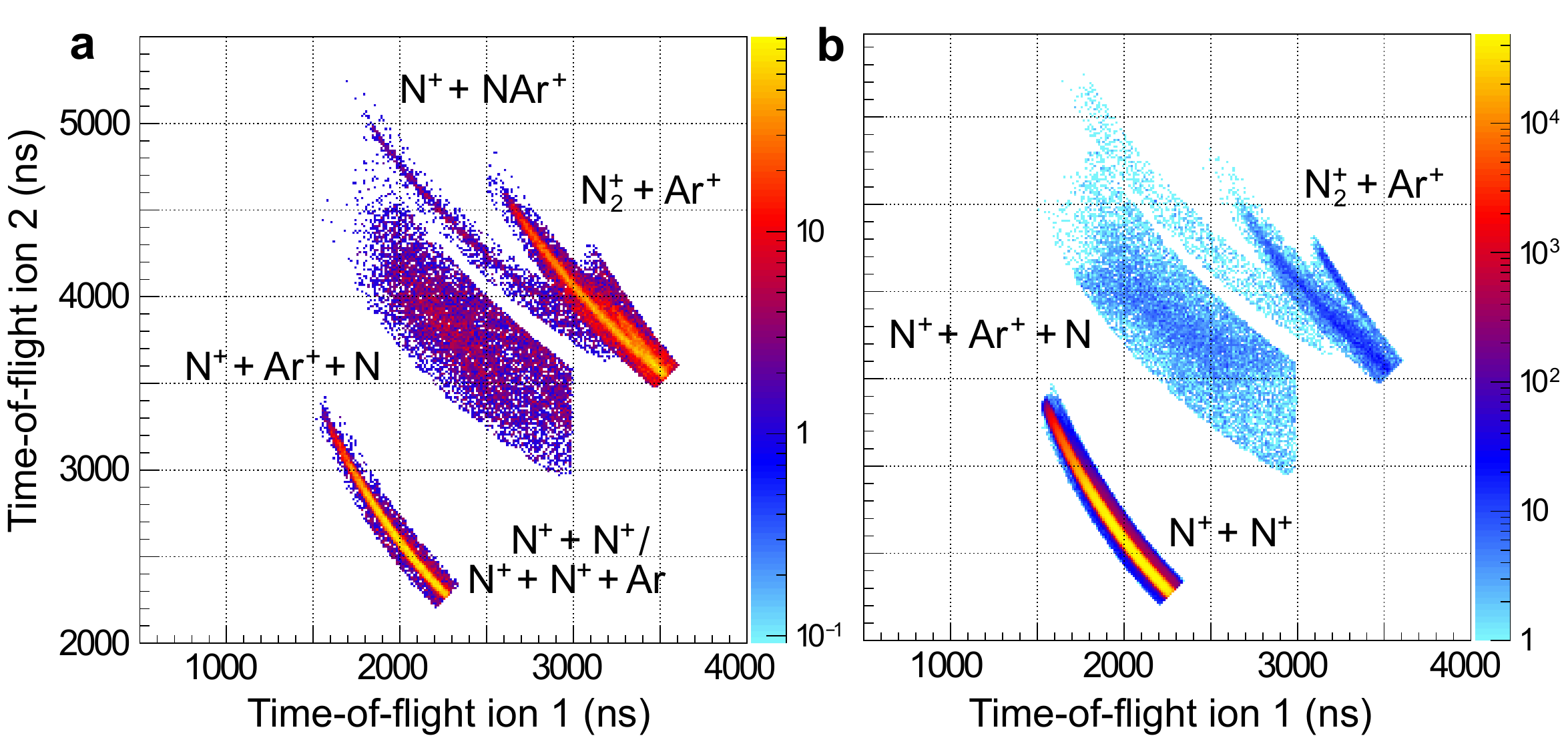}
\caption{\textbf{Measured three-body photoelectron-photoion-photoion coincidence spectrum.} Time-of-flight correlations of two ions detected in coincidence with a photoelectron from (\textbf{a}) Ar~2p photoionization at 260~eV and (\textbf{b}) N~1s photoionization at 420~eV. In panel (\textbf{a}), two additional fragmentation channels, N$^{+}$ + [NAr]$^{+}$ and N$^{+}$ + N$^{+}$ + Ar, appear alongside the channels observed in both panels: N$_2^{+}$ + Ar$^{+}$, N$^{+}$ + Ar$^{+}$ + N, and N$^{+}$ + N$^{+}$. The N$^{+}$ + N$^{+}$ + Ar and N$^{+}$ + N$^{+}$ channels overlap in panel (\textbf{a}); their separation and analysis are discussed in Section~\ref{subsec2}.}
\label{fig2}
\end{figure}

In Fig.~\ref{fig2} we present the time-of-flight correlation of two ions measured in coincidence with an electron. Panel (a) shows our results for a photon energy of $h\nu=260$~eV, while we used photons with an energy of 420 eV for the initial ionization of our gas target in Panel (b). In addition, we filtered both panels by requiring a photoelectron originating from the Ar and N$_2$ sites, respectively.
A Coulomb explosion involving a breakup into exactly two ionic fragments -- such as N$_2^{+}$ + Ar$^{+}$, N$^{+}$ + [NAr]$^{+}$ (and N$^{+}$~+~N$^{+}$ resulting from the ionization of uncondensed N$_2$ molecules) -- the fragments are emitted back-to-back (i.e., with momenta of equal magnitude but opposite direction) due to momentum conservation, resulting in a sum momentum of the fragments of zero. This yields narrow diagonal features in the coincidence maps shown in Fig.~\ref{fig2} and used to identify the ion pairs occurring in the final state. In contrast, the N$^{+}$~+~Ar$^{+}$~+~N channel appears as a broader structure, since the undetected neutral N atom carries away a part of that sum momentum. In addition, the dissociation channel N$^{+}$ + N$^{+}$ + Ar is expected to overlap with the aforementioned N$^{+}$ + N$^{+}$ channel, but|as we show below|can be disentangled by our coincidence measurement.


We observe a dimer fragmentation features belonging to all processes listed in Section~\ref{subsec2} and labeled them accordingly in Figs.~\ref{fig2}(a) and \ref{fig2}(b). In particular, we observe the breakup of the dimer after Ar 2p ionization into N$^{+}$ + [NAr]$^{+}$, i.e., events belonging to channel IV that requires a double charge transfer and an ion transfer.

\subsection*{Single Charge Transfer Pathways}\label{subsec1}

\begin{figure}[h]
\centering
\includegraphics[width=1.0\textwidth]{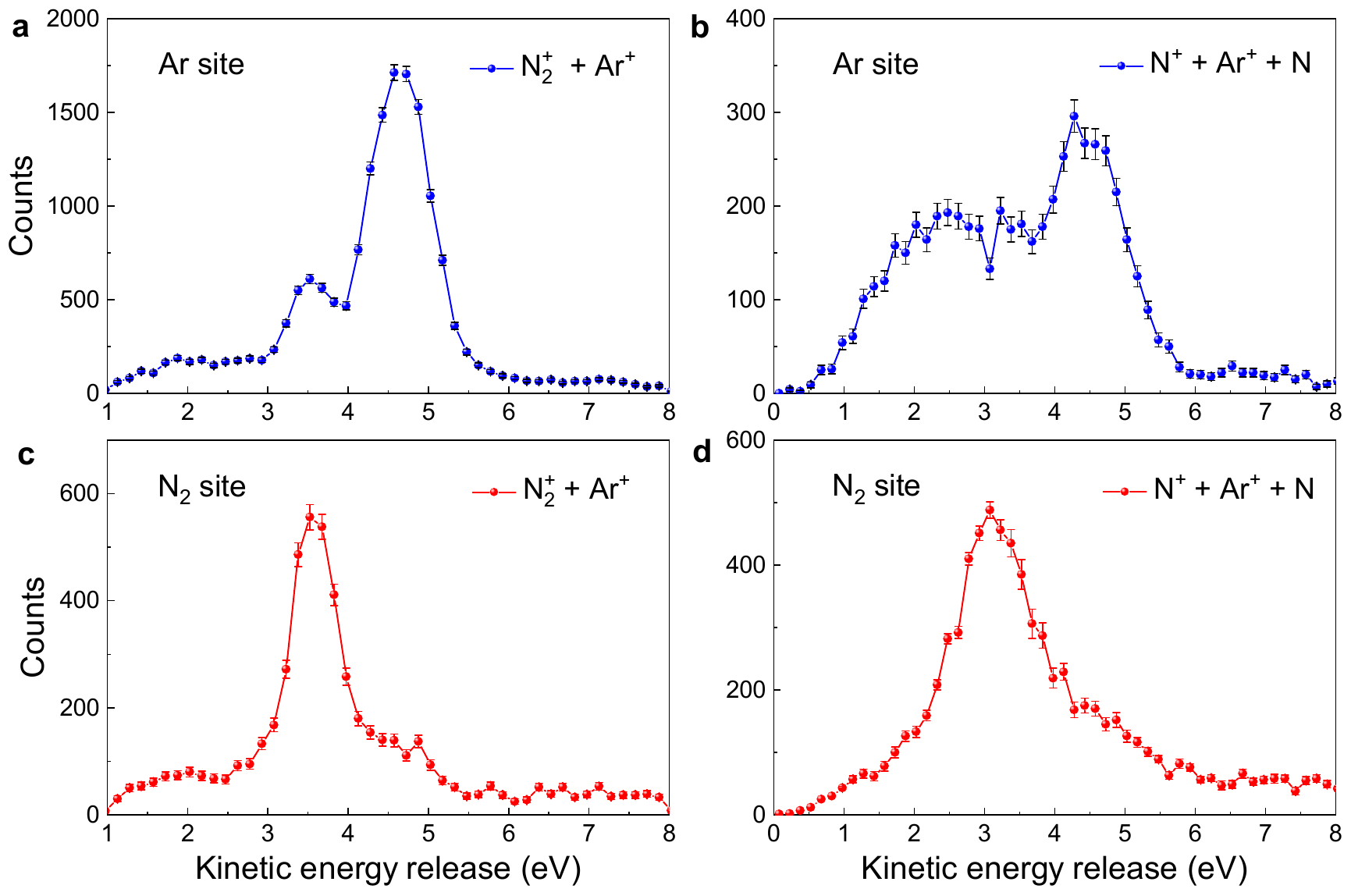}
\caption{\textbf{Kinetic energy release distributions of the fragment ions.} Measured KER spectra of two charged ions measured in coincidence with a photoelectron after breakup of the N$_2$Ar dimers into (\textbf{a}) N$_2^{+}$ + Ar$^{+}$ and (\textbf{b}) N$^{+}$ + Ar$^{+}$ + N following Ar~2p photoionization at 260~eV, and into (\textbf{c}) N$_2^{+}$ + Ar$^{+}$ and (\textbf{d}) N$^{+}$ + Ar$^{+}$ + N following N~1s photoionization at 420~eV. Error bars indicate statistical uncertainties.}\label{fig3}
\end{figure}

To elucidate the SCT mechanisms, we analyzed the kinetic energy release (KER) spectra of the ionic fragments. For the N$_2^{+}$ + Ar$^{+}$ channel [Ar site ionization, Fig.~\ref{fig3}(a)], the KER distribution features a dominant peak at 4.7~eV and a weaker peak at 3.5~eV. Using the point-charge Coulomb approximation, a KER of 3.5~eV corresponds to an internuclear distance R~=~\SI{4.11}{\angstrom}, consistent with the equilibrium distance of R~=~\SI{3.88}{\angstrom} \cite{10.1063/1.4871205}.
The 4.7~eV peak corresponds to R~=~\SI{3.06}{\angstrom}, indicating that N$_2$Ar$^{2+}$ ions are produced in an attractive region of the PES, and SCT occurs at a significantly shorter distance \cite{PhysRevA.104.042802}. This suggests the CT timescale is comparable to, or slower than, the cluster's vibrational motion.
The N$_2^{+}$ fragment can also be in a dissociative state, yielding the N$^{+}$ + Ar$^{+}$ + N channel. The corresponding KER distribution [Fig.~\ref{fig3}(b)] exhibits peaks at 4.5~eV and 2.4~eV. The higher-energy peak reflects direct N$_2^+$ fragmentation following SCT, where the total KER closely matches the main N$_2^+$ + Ar$^+$ peak. The lower-energy, broader peak arises when N$_2^+$ is repelled by Ar$^+$ before dissociation, and the neutral N atom carries away part of the energy.


To test whether the CT mechanism persists in more complex scenarios, we selectively ionized the N$_2$ site in N$_2$Ar dimers using photons of 420~eV. Following N~1s photoionization, Auger decay produces doubly charged N$_2^{2+}$Ar ions. The measured KER for the N$_2^{+}$ + Ar$^{+}$ channel [Fig.~\ref{fig3}(c)] shows a peak at 3.5~eV, consistent with the equilibrium separation. 
The N$^{+}$ + Ar$^{+}$ KER distribution [Fig.~\ref{fig3}(d)] shows lower values and broader widths compared with the N$_2^{+}$ + Ar$^{+}$ channel. This broadening occurs because N$_2^{2+}$ may begin to dissociate and release energy \textit{before} the SCT event.

\subsection*{Double Charge Transfer Pathways}\label{subsec2}

\begin{figure}[h]
\centering
\includegraphics[width=1.0\textwidth]{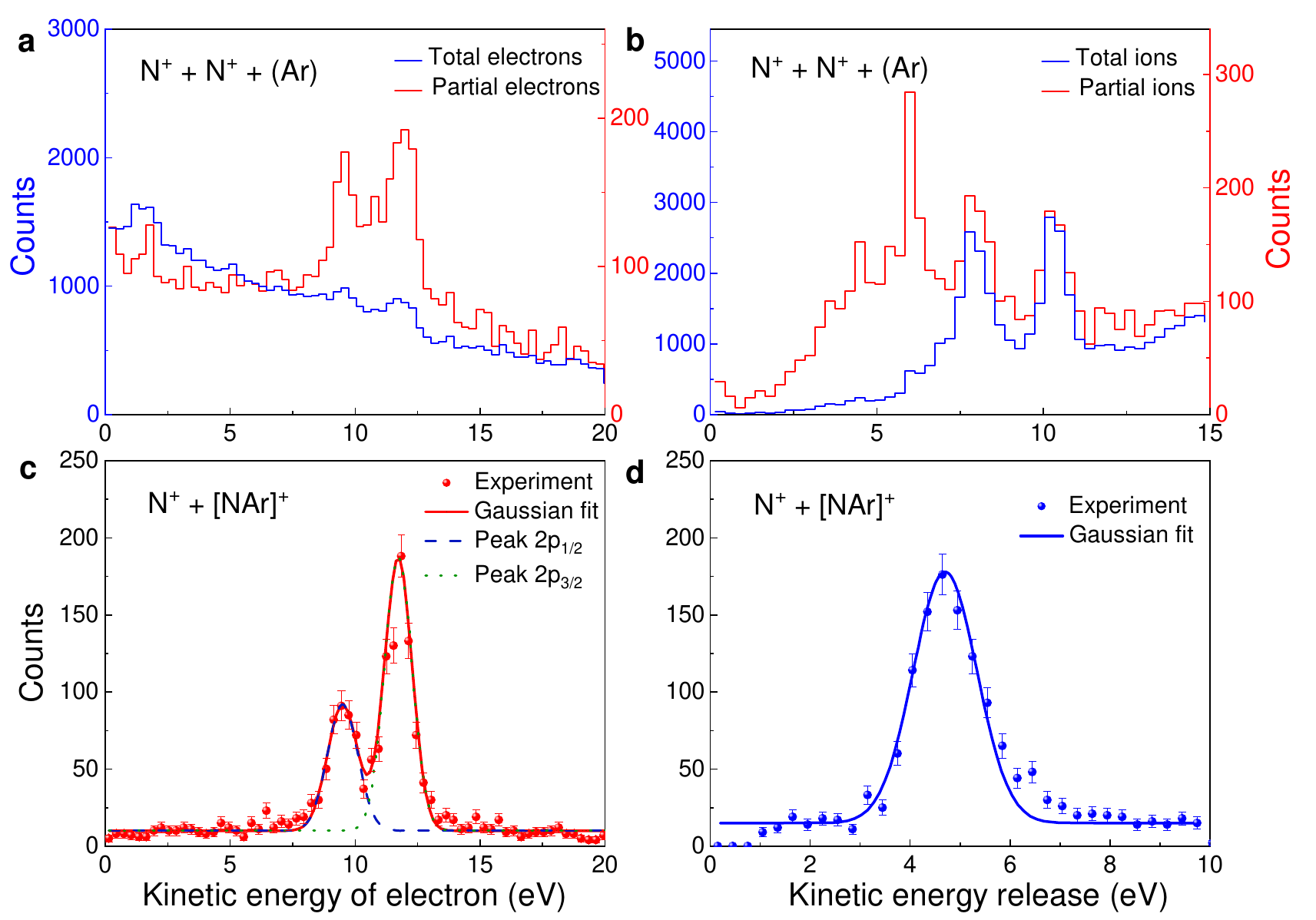}
\caption{\textbf{Kinetic energy distributions of the fragment ions and electrons.} (\textbf{a}) Measured electron kinetic energy and (\textbf{b}) KER spectra for N$^{+}$ + N$^{+}$ pairs following Ar~2p or N$_2$ valence DI at 260~eV. Shown are total events (blue line) and partial events (red line) selected by applying the momentum-conservation filter 5~a.u. $< |\sum p^i_z| <$ 20~a.u., which effectively suppresses most background contributions from N$_2$ monomer dissociation. The left and right $y$ axes indicate event counts for total and partial electrons in (\textbf{a}) and ions in (\textbf{b}), respectively. (\textbf{c}) Measured electron kinetic energy and (\textbf{d}) KER spectra for N$_2$Ar dimer breakup into N$^{+}$ + [NAr]$^{+}$ following Ar~2p photoionization at 260~eV. The two peaks in (\textbf{c}) correspond to photoelectrons from the Ar~2p$_{1/2}$ and Ar~2p$_{3/2}$ shells, respectively. Error bars represent statistical uncertainties.}\label{fig4}
\end{figure}

For the N$^{+}$ + N$^{+}$ channel, one might initially assume that it arises from dissociation of N$_2^{2+}$ produced by direct DI with a single 260~eV photon. However, the N$^{+}$ + N$^{+}$ channel accompanied by a neutral Ar atom can instead be generated via DCT, as confirmed by our calculated PESs in Fig.~\ref{fig1}. This assignment is further supported by our coincidence measurements with photoelectrons from Ar~2p inner-shell ionization, shown in Fig.~\ref{fig4}(a). The solid blue line represents the total electron spectrum from all events contributing to the N$^{+}$ + N$^{+}$ channel, while the solid red line shows a partial electron spectrum obtained by applying momentum-conservation conditions to exclude events with $|\sum p^i_z| \leq 5$~a.u. and $|\sum p^i_z| \geq 20$~a.u., where $p^i_z$ denotes the momentum of the $i$-th ion along the time-of-flight axis. The two distinct peaks observed in the red spectrum|absent in the blue|correspond to photoionization of the Ar~2p$_{3/2}$ and 2p$_{1/2}$ subshells \cite{KATO200739}. This provides direct evidence that the N$^{+}$ + N$^{+}$ channel originates from DCT processes in N$_2$Ar$^{2+}$ ions, rather than from N$_2^{2+}$Ar.


To further substantiate the presence of DCT processes, we compared the KER distribution of the N$^{+}$ + N$^{+}$ + Ar channel with that of the N$^{+}$ + N$^{+}$ channel, as shown in Fig.~\ref{fig4}(b). The solid blue and red lines correspond to the KER spectra of all N$^{+}$ + N$^{+}$ ion pairs and of the partial ion pairs selected using the same momentum-conservation conditions as in Fig.~\ref{fig4}(a), respectively. The two dominant peaks in the total-ion spectrum originate from the electronic states $\mathrm{D^3}\,\Pi_g$ and $\mathrm{D^1}\,\Sigma_u^+$ of N$_2^{2+}$ at 7.7~eV and 10.3~eV, which are assigned to the dissociation channels producing N$^{+}$($\mathrm{^3P}$)~+~N$^{+}$($\mathrm{^3P}$) and N$^{+}$($\mathrm{^3P}$)~+~N$^{+}$($\mathrm{^1D}$), respectively \cite{Lundqvist_1996}. In addition to these two peaks, the partial-ion spectrum exhibits a distinct feature near 6~eV with a broad 3-7~eV envelope. This structure is attributed to the lower excited states of N$_2^{2+}$, such as $\mathrm{a^3}\,\Pi_u$ and $\mathrm{A^1}\,\Pi_u$. Normally, dissociation of these states is difficult to detect because of their relatively long lifetimes. However, through DCT processes they are populated at bond lengths already beyond the potential barrier, enabling prompt dissociation. The emergence of this new peak therefore provides unambiguous evidence that the partial fragments of N$^{+}$ + N$^{+}$ originate from N$_2$Ar$^{2+}$ via DCT, which agrees well with our simulated PESs.

Interestingly, an exotic heavy-ion transfer channel, N$^{+}$ + [NAr]$^{+}$, is observed in site-specific X-ray-produced N$_2$Ar$^{2+}$ ions, as confirmed by the corresponding electron energy spectrum shown in Fig.~\ref{fig4}(c). To unravel the reaction mechanism, we revisit the PESs of N$_2$Ar$^{2+}$ shown in Fig.~\ref{fig1}. Selective inner-shell ionization of the Ar atom efficiently populates the $\mathrm{N}_2(\mathrm{X^1}\,\Sigma_g^+)$$\mathrm{Ar}^{2+}$($\mathrm{^3P}$) state. From there, the N–N vibrational wave packet evolves via a conical intersection into $\mathrm{N}_2^+(\mathrm{D}^2\Pi_g)$$\mathrm{Ar}^+(\mathrm{^2P})$. As the wave packet proceeds further, the PESs of $\mathrm{N}_2^+(\mathrm{D}^2\Pi_g)$$\mathrm{Ar}^+(\mathrm{^2P})$ and $\mathrm{N}_2^{2+}(\mathrm{a^3}\,\Pi_u)$$\mathrm{Ar}$ intersect at another conical intersection. Consequently, $\mathrm{N}_2^{2+}(\mathrm{a^3}\,\Pi_u)$$\mathrm{Ar}$ is populated from the initial Auger final state $\mathrm{N}_2(\mathrm{X^1}\,\Sigma_g^+)$$\mathrm{Ar}^{2+}$($\mathrm{^3P}$) through two sequential CT steps, i.e., a DCT process. The resulting N$_2^{2+}$Ar ions dissociate into N$^{+}$ + [NAr]$^{+}$, driven by Coulomb repulsion between the fragments. The measured KER spectrum [Fig.~\ref{fig4}(d)] exhibits a peak at 4.7~eV. During [NAr]$^{+}$ formation, all three atoms undergo substantial rearrangement; thus, the KER is not just given by Coulomb repulsion but also includes the vibrational energy of the [NAr]$^{+}$ product. Because the DCT process does not involve intrinsic energy loss, the accessible KER range can be derived from the energy difference between the initial and final states. For reactions starting from $\mathrm{N}_2(\mathrm{X^1}\,\Sigma_g^+)$$\mathrm{Ar}^{2+}$($\mathrm{^3P}$), this difference is approximately 6.45~eV when [NAr]$^{+}$ is in its vibrational ground state, suggesting that the observed [NAr]$^{+}$ carries 1.75~eV of vibrational excitation. The branching ratio of the $\mathrm{N}^{+} + [\mathrm{NAr}]^{+}$ channel reaches approximately 8\% relative to the dominant $\mathrm{N}_{2}^{+} + \mathrm{Ar}^{+}$ channel, highlighting a significant probability of heavy-ion transfer even in a simple dimer system.

Other than for ionization at the Ar site, no DCT processes are observed following ionization at the N$_2$ site. A likely explanation is that N$_2^{2+}$ ions produced via Auger decay after N~1s ionization are predominantly short-lived states that undergo prompt dissociation driven by Coulomb repulsion \cite{Püttner_2008}. For most of the final Auger states, the dissociation time scale of N$_2^{2+}$ is shorter than the SCT timescale, thus suppressing the SCT process. This accounts for the markedly lower SCT signal in Figs.~\ref{fig3}(c) and \ref{fig3}(d) compared with Ar site ionization in Figs.~\ref{fig3}(a) and \ref{fig3}(b), even though the N$_2$ site measurements were three times longer to compensate for differences in photoionization cross sections (N~1s at 420~eV vs. Ar~2p at 260~eV), yielding comparable total ionization counts. Likewise, the formation of [NAr]$^+$ is not observed, since N$_2^{+}$(1s$^{-1}$) rarely produces ground-state or low-lying excited N$_2^{2+}$ via Auger decay. Instead, highly excited N$_2^{2+}$ fragments into energetic N$^{+}$ ions, which are unlikely to be captured by Ar. These findings indicate that rapid fragmentation of N$_2$ can effectively suppress CT, although structural rearrangements within N$_2$ could in principle facilitate multiple CT pathways.

One might argue that N$_2^{2+}$Ar could be produced directly via one-photon DI of the N$_2$ site in N$_2$Ar dimers, subsequently leading to N$^{+}$ + [NAr]$^{+}$. To test this possibility, we performed an experiment with a photon energy of 43.1~eV, which lies above the DI threshold of N$_2$ (42.8~eV), but below the DI threshold of Ar (43.37~eV). The results clearly show that the N$^{+}$ + [NAr]$^{+}$ channel is absent, ruling out any contribution from one-photon DI of the N$_2$ site at 260~eV as well. In contrast to the previously proposed heavy-ion tunneling scenario from N$_2^{2+}$Ar \cite{article_NAr}, our findings demonstrate that [NAr]$^{+}$ formation proceeds via the DCT pathway described in expression IV, where electrons are transferred back and forth between the Ar and N$_2$ sites through two successive nonadiabatic transitions at conical intersections.




In conclusion, we investigated ultrafast multiple charge transfer (CT) processes mediated by nonadiabatic electron-nuclear dynamics in a weakly bound molecular system using multi-coincidence spectroscopy in combination with {\it ab initio} calculations. The N$_2$Ar dimer, which combines covalent and van der Waals bonding, serves as a well-defined prototype that is sufficiently complex to support sequential CT processes while remaining amenable to detailed experimental and theoretical analysis. Site-selective ionization of either the N$_2$ molecule or the Ar atom initiates intermolecular relaxation dynamics from well-defined initial charge states. Both single and double CT pathways were identified experimentally and supported by theoretical modeling, starting from the dicationic states N$_2$Ar$^{2+}$ and N$_2$$^{2+}$Ar. Double CT processes were unambiguously established by observing the N$^{+}$~+~N$^{+}$~+~Ar and N$^{+}$~+~[NAr]$^+$ fragmentation channels in coincident electron-ion measurements. In particular, the formation of [NAr]$^+$ was shown to result from a sequential double CT mechanism involving two nonadiabatic transitions at conical intersections, rather than from direct dissociation or heavy-ion tunneling. This mechanism leads to an effective ion-transfer process that qualitatively differs from the single CT dynamics typically observed in simpler atom-atom dimers. Our results demonstrate that sequential multiple CT events, coupled to ultrafast structural rearrangement, can drive complex reaction pathways even in a minimal mixed-interaction system. At the same time, the absence of double CT following ionization at the N$_2$ site highlights how rapid molecular dissociation can suppress CT dynamics. More generally, these findings illustrate how controlled model systems can be used to disentangle nonadiabatic charge redistribution mechanisms and to provide insight into CT-driven reaction dynamics that are relevant to a broad range of molecular and condensed-phase environments.

\section*{Methods}\label{sec11}

\backmatter

\bmhead{Experimental methods}

The experiments were performed using two synchrotron light sources at different photon energies: 260~eV and 420~eV were employed at the soft X-ray beamline P04 of the PETRA~III synchrotron (DESY, Hamburg, Germany) in timing mode (40 bunches, 192~ns bunch separation) with circularly polarized light \cite{VIEFHAUS2013151}; meanwhile, a photon energy of 43.1~eV was used at the Flexible PhotoElectron Spectroscopy (FlexPES) beamline of MAX~IV (Lund, Sweden) in single-bunch mode (320~ns bunch separation) with linearly polarized light \cite{Preobrajenski:ok5091}. In both facilities, a supersonic gas jet was formed, collimated by two skimmers, and crossed with the photon beam inside a COLTRIMS reaction microscope \cite{DORNER200095, Ullrich_2003, JAHNKE2004229}, with N$_2$Ar dimers produced by expanding a 50/50 gas mixture through a 30~$\mu$m nozzle at 11~bar (PIPE endstation PETRA~III) or a 50~$\mu$m nozzle at 5~bar (ICE endstation MAX~IV), both at 300~K. The employed COLTRIMS spectrometers consisted of an ion arm of 31~mm (PIPE endstation) or 34~mm (ICE endstation) acceleration length and an electron arm of 70~mm acceleration and 153~mm drift length (Wiley-McLaren time-of-flight focusing geometry, PIPE endstation) or 419~mm acceleration length (ICE endstation). All were equipped with microchannel plate detectors (active area of 120~mm diameter for the electron detector at PETRA~III and of 80~mm diameter for all other detectors) with hexagonal delay-line position readout \cite{JAGUTZKI2002244, JAGUTZKI2002256}. Electrons and ions were guided to position-sensitive microchannel plate detectors with delay-line readout using optimized electric and magnetic fields, 20.8~V/cm and 5.6~G at PETRA~III, and 12.4~V/cm and 7.7~G at MAX~IV, allowing for the complete 4$\pi$ collection of electrons with kinetic energies up to 30~eV and 23~eV, respectively. Data were recorded in list mode files, measuring the positions and times of flight of all reaction products for each ionization event; this enabled the extraction of the weak dimer signals and the reconstruction of particle momenta in offline analysis, despite most events originating from monomer ionization, and charge transfer channels were ultimately identified by selecting events with two ions of equal and opposite momentum, whose sharp back-to-back emission line in the PIPICO spectrum provides a unique signature of the final Coulomb explosion step.

\bmhead{Theoretical calculations}

The potential energy surfaces (PESs) of the linear [N$_2$Ar]$^{2+}$ along the N–N bond were initially calculated using a two-body potential model \cite{PhysRevB.80.165203}. This model incorporates only two-body correlation interactions, based on the known PESs of N$_2$, N$_2^{2+}$ and [NAr]$^{+}$, and neglects the three-body correlation. This approximation is justified for linear N$_2$Ar because one N atom is far from the Ar atom, rendering the three-body interaction negligible. To ensure higher precision, we subsequently performed \textit{ab initio} calculations of the PESs of linear [N$_2$Ar]$^{2+}$, N$_2$, N$_2^{2+}$, and [NAr]$^{+}$. These calculations utilized the multi-reference double excitation configuration interaction (MRD-CI) method \cite{10.1063/1.470544} with molecular orbitals optimized via the complete active space self-consistent field (CASSCF) method \cite{https://doi.org/10.1002/wcms.82, MOLPRO, 10.1063/5.0005081}. The active space comprised all valence orbitals, and approximately 1,000 reference configuration state functions were included in the CI calculations. The orbital wave functions were expanded in the aug-cc-pVTZ basis set. The resulting potential energy curves for linear [N$_2$Ar]$^{2+}$ are provided in the Supplementary Information.

\backmatter

\bmhead{Supplementary Information}


The Supplementary Information contains the resulting potential energy curves for the linear [N$_2$Ar]$^{2+}$ configuration.


\bmhead{Acknowledgements}
This work was supported by the National Key Research and Development Program of China (No. 2022YFA1604302), the National Natural Science Foundation of China (No. 12334011, No. 12541505, No. 12204498, No. 12474259, No. 12374262, and No. 12450404), and the Natural Science Foundation of Henan (No. 252300421304). F.T. acknowledges funding by the Deutsche Forschungsgemeinschaft (DFG, German Research Foundation) -- Project 509471550, Emmy Noether Programme. 

We acknowledge DESY (Hamburg, Germany), a member of the Helmholtz Association HGF, for the provision of experimental facilities. Parts of this research were carried out at PETRA~III. We would like to thank Jörn Seltmann and Moritz Hoesch for assistance during the experiments. Beamtime was allocated for proposal I-20240903. We acknowledge the MAX~IV Laboratory for beamtime on the FlexPES beamline under proposal 20241221. Research conducted at MAX~IV, a Swedish national user facility, is supported by Vetenskapsrådet (Swedish Research Council, VR) under contract 2018-07152, Vinnova (Swedish Governmental Agency for Innovation Systems) under contract 2018-04969 and Formas under contract 2019-02496. We also acknowledge support from the SXFEL and SHINE projects.


\section*{Declarations}

\begin{itemize}

\item Conflict of interest/Competing interests:

The authors declare no competing interests.

\item Data availability:

The data that support the plots within this article are available from the corresponding authors upon reasonable request.

\item Code availability:

The code that supports the plots within this article is available from the corresponding authors upon reasonable request.

\item Author contributions:

R.Dö., K.U., Y.W., F.T., T.J., and Y.J. supervised the project. R.Do., O.D.M., X.W., Y.Z., A.P., A.T., H.M., J.F., J.W., L.S., M.S., N.M., N.K., S.M.J., Z.S., N.W., K.U., F.T., T.J., and Y.J. performed the experiments. X.H., A.C., J.W., and Y.W. performed the simulations. R.Do., K.U., F.T., T.J., and Y.J. analyzed the data. R.Do., X.H., F.T., T.J., and Y.J. wrote the manuscript after in-depth discussions with K.U. and R.Dö. and with input from all the authors. 

\end{itemize}


\bibliography{sn-bibliography}

\end{document}